\documentclass[english]{elsarticle}
\usepackage[T1]{fontenc}
\usepackage[latin9]{inputenc}
\usepackage{geometry}
\usepackage{multirow}
\usepackage{amsmath}
\usepackage{amssymb}
\usepackage{graphicx}
\usepackage{esint}

\makeatletter

\AtBeginDocument{\providecommand\tabref[1]{\ref{tab:#1}}}
\AtBeginDocument{\providecommand\figref[1]{\ref{fig:#1}}}
\AtBeginDocument{\providecommand\subref[1]{\ref{sub:#1}}}
\providecommand{\tabularnewline}{\\}

\makeatother

\usepackage{babel}
\begin{document}

\title{Nonlinear Deconvolution by Sampling Biophysically Plausible Hemodynamic
Models}

\author{H.C. Ruiz Euler, J.P. Rebelo Ferreira Marques, H.J. Kappen}
\begin{abstract}
Non-invasive methods to measure brain activity are important to understand
cognitive processes in the human brain. A prominent example is functional
magnetic resonance imaging (fMRI), which is a noisy measurement of
a delayed signal that depends non-linearly on the neuronal activity
through the neurovascular coupling. These characteristics make the
inference of neuronal activity from fMRI a difficult but important
step in fMRI studies that require information at the neuronal level.
In this article, we address this inference problem using a Bayesian
approach where we model the latent neural activity as a stochastic
process and assume that the observed BOLD signal results from a realistic
physiological (Balloon) model. We apply a recently developed smoothing
method called APIS to efficiently sample the posterior given single
event fMRI time series. To infer neuronal signals with high likelihood
for multiple time series efficiently, a modification of the original
algorithm is introduced. We demonstrate that our adaptive procedure
is able to compensate the lacking of inputs in the model to infer
the neuronal activity and that it outperforms dramatically the standard
bootstrap particle filter-smoother in this setting. This makes the
proposed procedure specially attractive to deconvolve resting state
fMRI data. To validate the method, we evaluate the quality of the
signals inferred using the timing information contained in them. APIS
obtains reliable event timing estimates based on fMRI data gathered
during a reaction time experiment with short stimuli. Hence, we show
for the first time that one can obtain accurate absolute timing of
neuronal activity by reconstructing the latent neural signal. 
\end{abstract}

\ead{hruiz@science.ru.nl }
\maketitle

\section{Introduction}

Functional magnetic resonance imaging (fMRI) is an important method
to investigate sensory, motor and cognitive functions of the human
brain. This non-invasive technique to measure brain activity provides
indirect signals of the underlying neuronal dynamics because it measures
hemodynamic responses that represent changes in blood flow and oxygenation
levels. 

Inverting the non-linear hemodynamic system is important to understand
aspects of cognitive functions and for connectivity analysis at the
neuronal level. Nevertheless, reconstructing the brain activity from
the fMRI signal presents many challenges. For instance, the measured
signals might have components from non-neuronal hemodynamic sources
that can affect causal connectivity estimates \citep{david2008identifying}.
Moreover, the general inverse problem, also known as blind deconvolution,
is ill-posed because both the true parameters of the hemodynamic system
and the latent inputs are unknown. Therefore, given the fMRI time
course, the reconstruction of the neuronal signal is not unique. The
reason for this is that the hemodynamic parameters determine the delay
of the BOLD signal with respect to the underlying neuronal activity
giving a continuum of possible solutions. 

Activation studies employing general linear models (GLM) circumvent
this problem by using multiple repetitions of the same stimulus and
a canonical hemodynamic response function (HRF) together with time
and dispersion derivatives or basis functions that correct the HRF
\citep{friston1998event,friston1998nonlinear}. In these cases, temporal
information and the exact shape of the HRF are less likely to be crucial,
the models may not be very sensitive to temporal nuances in the data
and are usually not used to estimate the latent neuronal dynamics.

A very different approach to model the hemodynamic transformation
is to use physiologically informed models of BOLD responses, such
as the Balloon model \citep{buxton1998dynamics,friston2000nonlinear}.
Although this is a complex, highly parametrized biophysical model,
it is useful for jointly estimating the latent neuronal process and
the hemodynamic transformation, for instance with the well-known dynamic
causal modeling (DCM) \citep{friston2003dynamic} which assumes a
latent state space model (SSM). This deterministic method can be generalized
to a probabilistic method, i.e. stochastic DCM \citep{li2011generalised,daunizeau2009variational}.

The biological inspired models posses a rich HRF variability that
linear models do not capture. For instance, there is evidence of nonlinear
effects in the BOLD signal for high frequency stimulus presentations
\citep{birn2001spatial,wager2005accounting} and different stimuli
durations, e.g. when linear models are used, responses to $200\ ms$
stimuli are underestimated \citep{deneux2006using}. Therefore, non-linearities
in the hemodynamic system are important whenever we wish to estimate
the hidden neuronal activity and sophisticated fMRI analysis requires
methods that can handle these effects properly.

Estimating parameters of the biological plausible models involves
usually a maximum likelihood approach that requires the computation
of gradients with respect to the parameters. If these parameters affect
the noisy degrees of freedom, the computation of their gradient is
straight forward in the well-known Expectation-Maximization algorithm
(EM) for time series \citep{roweis2001learning}. For deterministic
degrees of freedom, one needs to resort to one of the following methods
to compute the gradient \citep{sengupta2014efficient}. The simplest
approach, currently used in DCM, is the finite difference method.
An alternative uses implicit differentiation of the error function
with respect to the relevant parameters. This gives a set of differential
equations that are integrated together with the forward model to compute
the gradient. This method\textendash{}called Forward Sensitivity (FS)
method\textendash{}was proposed in \citep{deneux2006using} in the
context of fMRI data. However, neither the reliability of this method
for fMRI data nor the resulting differential equations were discussed
in detail.  

The computation of the gradients to maximize the likelihood involves
the estimation of the hidden variables. This latent state estimation
is a problematic step where most of the methods differ. Until now,
there are roughly speaking three families of approaches for estimating
the latent process in the context of fMRI data. 

First, particle methods sample the dynamics in a forward and (possibly)
a backward pass to obtain Monte Carlo estimates of the posterior density.
As the number of samples (particles) grows, the estimates become more
accurate, however the computational cost of the backward pass makes
these methods unpractical in many cases \citep{briers2010smoothing,Douc2011,lindsten2013backward}.
Moreover, it has been shown that the method proposed here outperforms
the vanilla implementations of particle filter methods, namely the
bootstrap filter-smoother and the forward-backward smoother \citet{ruiz2017particle}.
In \citep{murray2011particle,johnston2008nonlinear}, methods of this
family were applied to estimate hidden neuronal and hemodynamic states
from fMRI data.

Second, variational Bayesian methods approximate the posterior distribution
with a variational density that maximizes a lower bound on the evidence.
This has been applied in the context of fMRI data in the dynamic expectation
maximization algorithm and its generalizations, which use models formulated
in terms of generalized coordinates of motion \citep{friston2008avariational,friston2010generalised,friston2008bvariational}.

Third, Gaussian methods approximate the posterior distributions by
Gaussians under the assumption that one can obtain a consistent minimum
variance estimator by recursive propagation and update of the mean
and variance of the target densities. Important examples are the extended
Kalman filter and the unscented Kalman filter. For fMRI data, \citep{riera2004state}
used the local linearization filter \citep{jimenez2003local}. More
recently, \citep{havlicek2011dynamic} proposed an estimation scheme
using cubature Kalman filtering in the forward pass and a cubature
Rauch-Tung-Striebel smoother in the backward pass.

Interestingly, the methods mentioned above \citep{friston2008bvariational,havlicek2011dynamic,riera2004state}
are the only available techniques applied to fMRI data shown capable
of estimating the inputs together with the hidden states. This is
important because the assumption of known exogenous inputs does not
always hold--for instance in resting state fMRI data there are no
stimuli present during the measurements--or it might be an implausible
assumption for higher cognitive areas \citep{lindquist2008statistical}.
However, it is not clear how accurate these methods are for different
stimulus durations and frequency, because either they need restrictive
parameterization of the inputs \citep{riera2004state}, or the experimental
design on which they were tested had stimulus durations of several
seconds and hence, their validity is restricted to inputs of the same
time scale as the BOLD signal \citep{sreenivasan2015nonparametric,friston2008bvariational}.

A precise measure of the timing of brain activity is important to
better understand the neural substrate of cognitive processes in the
brain. Detecting correct input timings to the brain regions from fMRI
measurements may help in our understanding of mental chronometry,
i.e. the sequential patterns of brain activity involved during cognitive
processes. A prominent example for the need of precise sequential
information is the reconstruction of causal interactions among brain
regions \citep{friston2003dynamic}. Although it has been shown that
small relative timings differences are detectable in a given location
\citep{miezin2000characterizing,liao2002estimating,formisano2003tracking,katwal2013measuring},
it is generally thought that accurate measurements of the absolute
timing of neuronal activity is not feasible. Indeed, due to the indirect
measurement procedure involved in fMRI and the ill-defined nature
of the blind deconvolution problem, this is an extremely challenging
problem.

To the best of our knowledge, there is no analysis available on the
precision of absolute input timing estimates from real fMRI data,
especially for very short stimuli in the order of hundreds of milliseconds.
Interesting work in this direction is found in \citep{havlicek2011dynamic}.
There, examples of estimations on synthetic data using short inputs
were obtained but there was no analysis on the precision of the input
timings recovered. Due to the lack of a ground truth that guides the
validation of methods when applied to real fMRI time series, we believe
that the timing estimates of short stimuli is an important aspect
when considering validating methods of fMRI deconvolution. Furthermore,
analyzing the errors in absolute timings is also an important tool
to asses the quality of the estimates for the hemodynamic parameters
that influence the delay in the response.

\subsection*{Outline}

In this article, we estimate the neuronal activity and absolute input
timings in individual regions of interest (ROIs) from single-event
fMRI time series gathered in a reaction time experiment with very
short stimuli ($150\ ms$). First, we analyze in detail the sensitivity
of the BOLD delay to changes in the hemodynamic parameters to identify
the most relevant parameters affecting the delay of the BOLD signal
relative to the input. We then derive the differential equations involved
in the estimation of these parameters via the FS method. Here, we
assume deterministic dynamics and that the input is known. In addition,
we analyze and evaluate this method using synthetic data and estimate
the relevant parameters for the fMRI time series to improve the timing
estimates of unknown inputs. 

Second, we show the feasibility of latent state estimation for fMRI
time series using APIS with excellent sampling efficiency. This is
non-trivial because of the complex (Balloon) observation model. In
\citep{ruiz2017particle} we showed that the APIS method significantly
outperforms particle filtering/smoothing methods on complex time series
inference tasks. We thus demonstrate that the advantage of APIS over
particle filtering methods, in terms of efficiency and accuracy, also
applies to fMRI time series data. In addition, during the input estimation
procedure we adapt the neuronal noise variance using EM. This is an
important and novel step to maximize the data likelihood for a large
amount of time series at once and to obtain estimates of the neuronal
activity with large amplitudes and accurate event timings.

To validate our method, we compare the estimated timing errors with
errors obtained when synthetic data is used, i.e. when the ground
truth model is known. We show that, given sufficiently accurate estimates
of the hemodynamic parameters, the proposed method can infer the absolute
input timing of single events reliably by reconstructing the hidden
neuronal activity.

\section{Method}

\subsection{fMRI Data}

We analyze fMRI time series obtained from the experimental setting
in \citep{narsude2015three}, consisting of subjects reacting to the
occurrence of either a visual stimulus (V), an auditory stimulus (A)
or a simultaneous combination of both stimuli (AV). Whenever the subjects
perceived the stimulus, they had to press a button as fast as possible.
The modality of the stimulus presented was random and had a duration
of $150\ ms$. We denote the presentation of a stimulus or the reaction
as an \textquotedbl{}event\textquotedbl{} in the sensory or motor
ROIs respectively. 

We consider only the first subject of this study, who participated
in two trials. Each trial consisted of 30 stimuli presented at a rate
of 1 stimulus per $16\ s$ (with some jitter of 0.2 seconds). The
response times were recorded and the temporal resolution of the acquisition
was $TR=0.4\ s$. Because this work focuses on single trial measurements,
the data is denoised using manual independent component classification
as proposed in \citep{griffanti2017hand}. A standard GLM analysis
is then used to delineate the motor, auditory and visual cortex used
in this task. Regions of interest of about 500 voxels ($2mm$ isotropic)
are defined for the audio (A ROI), visual (V ROI) and left motor cortices
(M ROI) and time series extracted from them. The time series of subjects
2 and 3 are too noisy when performing a standard GLM analysis and
have to be discarded.

We estimate the neuronal activity around each event separately. For
this, the original time course is divided in 30 segments of length
$T=16\ s$ around the event time. This amounts to 41 observations
in each time series (the first defining $t=0$) and the event time
falls around $3.2$ seconds.

To characterize the relative BOLD change $\left\{ \hat{y}_{i}\right\} _{i=1,\dots,41}$,
each time course $\left\{ y_{i}\right\} _{i=1,\dots,41}$ is centralized
and normalized by the mean $\mu_{y}$ of its \textquotedbl{}null\textquotedbl{}
distribution,
\[
\hat{y}_{i}=\frac{y_{i}-\mu_{y}}{\mu_{y}}.
\]

The \textquotedbl{}null\textquotedbl{} distribution of each time series
is defined as the set of data points outside a time window after the
stimulus, when the neuronal activity is assumed to have the baseline
value ($z_{t}=0$). The window is chosen such that the variance of
the data points included is minimal and is estimated for each event
separately to account for long time scale fluctuations in the data.

\subsection{Modeling the BOLD Signal}

Similar to \citep{friston2000nonlinear}, we consider a single region
with a 5-dimensional dynamic state $x=(z,s,f,q,v)$. However, unless
stated otherwise, the neuronal activity $z$ follows stochastic dynamics
given by

\begin{equation}
dz=-A(z-u(z_{t},t))dt+\sqrt{A}\sigma_{z}dW_{z}\label{eq:neuronal-activity-dyn}
\end{equation}

The parameter $A$ sets the time scale of the neuronal response and
$dW_{z}\sim N(0,dt)$ is a Wiener process with variance%
\footnote{In all the simulations we use this discretization step.%
} $dt=0.01$. According to the literature, typical values observed
for the neuronal lags are around $5-35\ ms$ \citep{smith2011network}.
We set $A$ such that the system has a characteristic time scale of
$1/A=20\ ms$, but our results are robust against changes in $A$.
The term $\sqrt{A}$ in the noise ensures that the stationary distribution
remains invariant to the time scale. 

Notice that we define process (\ref{eq:neuronal-activity-dyn}) to
have an unknown input $u(z,t)$ that we wish to estimate. In general,
$u(z,t)$ can be any parametrized function \citep{kappen2016adaptive},
but in this paper it was chosen to have the simple form $u(z,t)=I_{z}(t)z+I_{t}(t)$
with $I_{z}(t)$ and $I_{t}(t)$ as time varying functions to be learned
\citep{ruiz2017particle}.

The neuronal activation $z_{t}$ is passed through the following nonlinear
deterministic transformation defining the dynamics of the other four
variables; two Hemodynamic equations \citep{friston2000nonlinear}
\begin{align}
ds & =\left(\epsilon z-\frac{s}{\tau_{s}}-\frac{f-1}{\tau_{f}}\right)dt\label{eq:hemodyn-eqs}\\
df & =sdt\nonumber 
\end{align}
 and two equations of the Balloon model \citep{buxton1998dynamics},

\begin{align}
dq & =\frac{1}{\tau_{0}}\left(f\frac{1-(1-E_{0})^{1/f}}{E_{0}}-v^{(1/\alpha)-1}q\right)dt\label{eq:balloon-model}\\
dv & =\frac{1}{\tau_{0}}\left(f-v^{1/\alpha}\right)dt.\nonumber 
\end{align}

The BOLD signal change is given by

\begin{equation}
\hat{y}(t)=B(q_{t},v_{t}|\vartheta)+\sigma_{y}dW_{y}\label{eq:BOLD-signal}
\end{equation}
where $B(q_{t},v_{t}|\vartheta):=V_{0}\left[k_{1}\left(1-q_{t}\right)+k_{2}\left(1-\frac{q}{v}\right)+k_{3}\left(1-v\right)\right]$
and $\vartheta$ denotes all parameters of the system.

In addition, we consider as prior for the initial condition $x_{0}=(z_{0},s_{0},f_{0},q_{0},v_{0})$
a normal distribution%
\footnote{Notice that due to the small discretization step $dt$ and noise levels
used here, the log-transformation of the hemodynamic variables was
not required \citep{stephan2008nonlinear}. Nevertheless, it is straightforward
to use this transformation in our procedure.%
} with mean $\mu_{0}=(\mu_{z,0},\mu_{s,0},\mu_{f,0},\mu_{q,0},\mu_{v,0})$
and a covariance given by a diagonal matrix with small entries $\sigma_{z,s,f,q,v,0}^{2}$.

In this article, the parameters of the model were fixed based on \citep{havlicek2015physiologically},
with the exception of $\tau_{0}$, $\tau_{f}$ and the neuronal lag
$1/A$. These are given in table \tabref{Parameters-of-the-BOLD-trafo},
where the values marked with TBD are to be determined. The values
for $k_{1,2,3}$ correspond to those for a gradient echo (GE) sequence
at a field strength of $7\ T$ with an echo time of $TE=26\ ms$,
according to the settings in \citep{narsude2015three}.

\begin{table}
\centering{}%
\begin{minipage}[t]{1\columnwidth}%
\begin{center}
\begin{tabular}{|c|c|c|c|c|c|c|c|}
\hline 
Parameter & Value & Parameter & Value & Parameter & Value & Parameter & Value\tabularnewline
\hline 
\hline 
$A$ & 50 Hz & $\sigma_{z}$ & TBD & $\mu_{z,0}$ & $0$ & $\sigma_{z,0}$ & $\sigma_{z}/\sqrt{2}$\tabularnewline
\hline 
\hline 
$\epsilon$ & 0.8 & $E_{0}$ & 0.4 & $\alpha$ & 0.32 & $\sigma_{s,0}$ & TBD\tabularnewline
\hline 
$\tau_{s}$ & 1.54 & $k_{1}$ & 8.4 & $V_{0}$ & 0.04 & $\sigma_{f,q,v,0}$ & TBD\tabularnewline
\hline 
$\tau_{f}$ & TBD & $k_{2}$ & 0 & $\mu_{s,0}$ & $0$ & $\sigma_{y}$ & TBD\tabularnewline
\hline 
$\tau_{0}$ & TBD & $k_{3}$ & 1 & $\mu_{f,q,v,0}$ & $1$ &  & \tabularnewline
\hline 
\end{tabular}
\par\end{center}

\caption{Parameters for the neural dynamics (top row) and for the BOLD transformation
(bottom rows). \label{tab:Parameters-of-the-BOLD-trafo}}
\end{minipage}
\end{table}

\subsection{Estimating the Parameters of the Model }

We estimate the observation noise for each time series from its null
distribution. This gives for each case a slightly different noise
estimate around $\sigma_{y}\simeq0.002$. 

The variance of the initial state $x_{0}$ is set to the variance
of the stationary distribution induced by the Ornstein-Uhlenbeck process
for the uncontrolled dynamics in eq. (\ref{eq:neuronal-activity-dyn})
when $u(z,t)=0$. Hence, the variance for $p(z_{0})$ is $\sigma_{z,0}=\sigma_{z}/\sqrt{2}$
and all others are estimated by forward sampling of the system. Because
all other dynamic variables are deterministic, the stationary distribution
defining our prior over the process is determined by a single free
parameter $\sigma_{z}$, the noise of the neuronal activity. This
free parameter is adapted to infer inputs that maximize the likelihood.

\subsubsection{Forward Sensitivity Equations \label{sub:Forward-Sensitivity-Equations}}

The hemodynamic parameters are estimated using the Forward Sensitivity
method. The negative log-likelihood or \textquotedbl{}total cost\textquotedbl{}
$C(\vartheta)=\int_{0}^{T}c_{t}(\vartheta)dt$ is an implicit function
of the parameters $\vartheta$ in our model, where $c_{t}(\vartheta)=\sum_{t_{obs}}\delta(t-t_{obs})[\hat{y}_{t}-B(q_{t},v_{t}|\vartheta)]^{2}/2\sigma_{y}^{2}$
summing over the observation times $t_{obs}=t_{1},\dots,t_{41}$.
Thus, given an infinitesimal change $\delta\vartheta$ we can expressed
the change in $c_{t}(\vartheta)$ at each time point $t\in[0,T]$
by 
\[
\delta c_{t}=\frac{\partial c_{t}}{\partial q_{t}}\delta q_{t}+\frac{\partial c_{t}}{\partial v_{t}}\delta v_{t}+\sum_{i}\frac{\partial c_{t}}{\partial\vartheta_{i}}\delta\vartheta_{i}
\]
where $\delta q_{t},\delta v_{t}$ also depend on $\vartheta$ and
are obtained by the variation of the hemodynamic states $q_{t},v_{t}$
w.r.t. $\vartheta$. 

We only consider changes in $\tau_{f}$ and $\tau_{0}$ since these
parameters have a strong effect on the delay between peaks of the
neuronal activity and the BOLD signal, keeping all other parameters
fixed. However, the following analysis can be done for any other parameters
in the hemodynamic model.

The aim is to obtain the derivatives of $C(\tau_{f},\tau_{0})$ by
integrating a system of differential equations for \textquotedbl{}bar\textquotedbl{}
variables $\bar{q}_{0}(t)=dq_{t}/d\tau_{0},\ \bar{v}_{0}(t)=dv_{t}/d\tau_{0}$
and $\bar{q}_{f}(t)=dq_{t}/d\tau_{f},\ \bar{v}_{f}(t)=dv_{t}/d\tau_{f}$.
The total cost is used to update the parameters $\tau_{f,0}$ in a
gradient descent fashion. From the variation $\delta\tau_{0}$ we
obtain the total derivative of $dc_{t}/d\tau_{0}$,
\[
\frac{dc_{t}}{d\tau_{0}}=\frac{\partial c_{t}}{\partial q_{t}}\bar{q}_{0}(t)+\frac{\partial c_{t}}{\partial v_{t}}\bar{v}_{0}(t)
\]
where $\bar{q}_{0}(t),\bar{v}_{0}(t)$ follow the dynamic equations,
\begin{gather*}
\dot{\bar{q}}_{0}=-\frac{1}{\tau_{0}}\left(\dot{q}+v^{1/\alpha-1}\bar{q}_{0}+(\frac{1}{\alpha}-1)v^{1/\alpha-2}q\bar{v}_{0}\right)\\
\dot{\bar{v}}_{0}=-\frac{1}{\tau_{0}}\left(\dot{v}+\frac{1}{\alpha}v^{1/\alpha-1}\bar{v}_{0}\right)
\end{gather*}

The total derivative $dc_{t}/d\tau_{f}$ obtained by the variation
$\delta\tau_{f}$ is 
\[
\frac{dc_{t}}{d\tau_{f}}=\frac{\partial c_{t}}{\partial q_{t}}\bar{q}_{f}(t)+\frac{\partial c_{t}}{\partial v_{t}}\bar{v}_{f}(t)
\]
where $\bar{q}_{f}(t),\bar{v}_{f}(t)$ follow
\begin{gather*}
\dot{\bar{s}}_{f}=-\frac{1}{\tau_{s}}\bar{s}_{f}-\frac{1}{\tau_{f}}\left(\bar{f}_{f}+\frac{f-1}{\tau_{f}}\right)\\
\dot{\bar{f}}_{f}=\bar{s}_{f}\\
\dot{\bar{v}}_{f}=-\frac{1}{\tau_{0}}\left(\frac{1}{\alpha}v^{1/\alpha-1}\bar{v}_{f}-\bar{f}_{f}\right)\\
\dot{\bar{q}}_{f}=-\frac{1}{\tau_{0}}\left(v^{1/\alpha-1}\bar{q}_{f}+(\frac{1}{\alpha}-1)v^{1/\alpha-2}q\bar{v}_{f}-\bar{E}(f)\bar{f}_{f}\right)
\end{gather*}
with $\bar{E}(f):=E(f)+\ln(1-E_{0})\frac{(1-E_{0})^{1/f}}{E_{0}f}$.

Having the above extended system of differential equations, we can
initialize the bar variables to zero and integrate the entire system
forward in time. After integration on the interval $t\in[0,T]$, we
update the parameters using $dC/d\tau_{0}$ and $dC/d\tau_{f}$ respectively.
This is done iteratively until convergence of the parameters. 

Before applying this estimation procedure to real data, we validate
it on synthetic data generated by numerical integration of a deterministic
system ($\sigma_{z}=0$). As input to the system we use a box function
with $150\ ms$ length and an on-set time at 3.2 seconds. The amplitude
of the input is set to 1. The ground truth values of the parameters
are taken to be those in table \tabref{Parameters-of-the-BOLD-trafo}
together with $\tau_{0}=1.02$, $\tau_{f}=2.44$. To generate the
synthetic data from this model, a realistic scenario is considered
with $TR=0.4$ and a Gaussian measurement noise of $\sigma_{y}=0.002$.
For the estimation procedure, the parameters $\tau_{0}^{(0)},\tau_{f}^{(0)}$
are initialized randomly from a log-normal distribution s.t. its log-transformed
distribution has variance $0.6$ and mean $(\log\tau_{0}^{(0)},\log\tau_{f}^{(0)})=(0.4,0)$.
All other parameters of the system are fixed to the ground truth values.

After validation of the FS method on synthetic data, we fit the parameters
$\tau_{0},\tau_{f}$ to the mean fMRI time series of each ROI (A,V,M)
in trial 1 and assuming an input to that ROI. The resulting models
are used to estimate the input timing of individual time series in
trial 2 without any assumptions on the inputs.

Since the sensory ROI A and V do not respond or respond very weakly
to the stimulus modality V and A respectively, we do not consider
these events when estimating the parameters of the corresponding ROI.
Hence, for the auditory and visual regions, there are only 20 signals
available in each ROI. On the contrary, all stimuli elicit responses
in the motor ROI so we use all 30 events to compute the mean BOLD
response. The input to the model for all ROIs is also a box function.
For the A and V regions, the on-set time is the mean event time of
all the relevant time series. The on-set time of the input to the
motor ROI is considered to be the mid-time between the mean stimulus
and reaction times.

\subsection{Hidden State Estimation }

Given an fMRI time series $\hat{y}_{0:T}=\{\hat{y}_{0},\hat{y}_{t_{1}},\dots,\hat{y}_{T}\}$
we are interested in an efficient estimation of the posterior $p(z_{[0,T]}|\hat{y}_{0:T})$,
where the prior $p(z_{[0,T]})$ is given by (\ref{eq:neuronal-activity-dyn})
and $z_{[0,T]}$ denotes the continuous path or process from time
$t=0$ to $t=T$. The likelihood is given by the observation model
(\ref{eq:hemodyn-eqs})-(\ref{eq:BOLD-signal}).

The posterior or smoothing distribution can be seen as the solution
to a stochastic optimal control problem \citep{fleming1982optimal,kappen2012optimal}.
Using this relation we express the problem of estimating the posterior
over the neuronal activity as a Path Integral control problem \citep{kappen2005linear,kappen2011optimal}.

Recent results in \citep{ruiz2017particle} show that having a sufficiently
good parametrization of the control improves dramatically the efficiency
of sampling in terms of the effective sample size. This efficiency
is achieved with an adaptive importance sampling method called APIS.
The sampler learns iteratively a time dependent feedback controller
to adapt the process such that the likelihood of the samples increase.
This in turn reduces the variance of the importance weights and increases
the effective sample size.

Roughly, APIS works as follows. At each iteration, $N$ samples are
initialized with a Gaussian prior $p(x_{0})$. The samples are propagated
in time by integrating the (stochastic) dynamic equations (\ref{eq:neuronal-activity-dyn})-(\ref{eq:balloon-model}).
The sampled paths $\left\{ z_{[0,T]}^{i}\right\} _{i=1,\dots,N}$
are used to compute statistics over the posterior using the respective
importance weights obtained from the likelihood and a quadratic control
cost acting as a regularizer of the control function $u(z,t)$. Using
these statistics, the values $I_{z}(t),I_{t}(t)$ are updated at each
time point to estimate the control/input function used in the next
iteration. This is repeated until convergence of the effective sample
size. 

In summary, we use importance sampling adaptively to estimate the
hidden diffusion process by regarding the input $u(z,t)=I_{z}(t)z+I_{t}(t)$
as a control function to be estimated. A good approximation of $u(z,t)$
allows steering the samples such that the resulting BOLD signal has
high likelihood. We refer the interested reader to \citep{ruiz2017particle}
for further details.

\subsection{Noise Adaptation }

One of the features of path integral control theory is that the difference
between the posterior distribution relative to the prior process strongly
depends on the noise level $\sigma_{z}$. The control is proportional
to $\sigma_{z}$. Thus, if $\sigma_{z}$ is small, the control will
be small and the posterior deviates only slightly from the prior.
This affects directly the likelihood under a strong model mismatch.
For instance, if we do not model any input to the neuronal system,
the prior is the stationary distribution. Ideally, the controller
must adapt the system such that first, the efficiency of the sampler
increases and, second, the model mismatch is overcome. This requires
strong controllers capable of acting as input signals. However, it
is not clear a priori the amount of noise required to obtain sufficiently
strong inputs that give a high likelihood solution. 

We propose an approach that achieves an efficient sampling and high
likelihood by starting with a simpler low-noise problem and increase
the noise level by gradient ascent on the log-likelihood function
in the direction of the noise $\sigma_{z}$. We derive the update
rule for $\sigma_{z}$ using the EM approach.

In its general form, the EM algorithm maximizes iteratively the expected
log-likelihood of the complete data 
\[
\vartheta^{n+1}=\underset{\vartheta}{\text{argmax}}\mathbb{E}\left[\log p\left(x_{[0,T]},\hat{y}_{0:T}|\vartheta\right)|\hat{y}_{0:T},\vartheta^{n}\right].
\]
 The expectation is over the posterior $p(x_{[0,T]}|\hat{y}_{0:T},\vartheta^{n})$,
given by the solution of APIS for a fixed value of the parameters
in the $n$-th iteration. Due to the Markov property of the SSM, the
complete data log-likelihood separates into three terms involving
the logarithm of the prior over initial conditions, the likelihood
of the observations and the transition probability between two time
slices, all conditioned on the unknown parameters $\vartheta$. 

We only focus on the dynamic noise $\sigma_{z}$ found in the latter
term of the complete data log-likelihood. Hence, keeping all other
parameters fixed, we seek to maximize 
\[
Q_{z}(\sigma_{z},\sigma_{z}^{(n)})=\mathbb{E}\left[\sum_{t=dt}^{T}\log p\left(z_{t}|z_{t-dt},\sigma_{z}\right)|\hat{y}_{0:T},\sigma_{z}^{(n)}\right]
\]
w.r.t. $\sigma_{z}$, where the sum is over all discretization steps
on the interval $(0,T]$. In general, this transition probability
can be written as a Gaussian probability density for sufficiently
small $dt$, giving 
\[
\log p\left(z_{t}|z_{t-dt},\sigma_{z}\right)=-\log\sigma_{z}-\frac{[dz_{t}-f(z_{t},t)dt]^{2}}{2\sigma_{z}^{2}dt}+\text{const}
\]
where $f(z_{t},t)$ is the drift of the uncontrolled system, e.g.
here $f(z_{t},t)=-Az_{t}$. Thus, the gradient of $Q_{z}$ w.r.t.
$\sigma_{z}$ is
\[
\frac{\partial Q_{z}}{\partial\sigma_{z}}=-\frac{N_{T}}{\sigma_{z}}+\frac{1}{\sigma_{z}^{3}}\mathbb{E}\left[\sum_{t=dt}^{T}\frac{[dz_{t}-f(z_{t},t)dt]^{2}}{dt}|\hat{y}_{0:T},\sigma_{z}^{(n)}\right]
\]
where $N_{T}$ the number of discretization steps $N_{T}dt=T$. Using
(\ref{eq:neuronal-activity-dyn}), notice that when sampling with
the control function $u(z_{t},t)$ and noise $\sigma_{z}^{(n)}$ we
have $[dz_{t}-f(z_{t},t)dt]^{2}=\left[Au(z_{t},t)dt+\sqrt{A}\sigma_{z}^{(n)}dW_{t}\right]^{2}$. 

Although the EM algorithm allows for a single step update of the variance
$\sigma_{z}^{2}$, the procedure is more stable if we use gradient
ascent on $\sigma_{z}$ whenever the effective sample size is above
a pre-defined threshold $\gamma_{\sigma}$,
\begin{equation}
\sigma_{z}^{(n+1)}=\sigma_{z}^{(n)}+\eta\frac{\Sigma^{(n)}-1}{\sigma_{z}^{(n)}}\label{eq:noise-update-rule}
\end{equation}
where $\eta$ is the learning rate and 
\[
\Sigma^{(n)}=\frac{1}{T}\mathbb{E}\left[\sum_{t=dt}^{T}\left[\frac{Au(z_{t},t)dt+\sqrt{A}\sigma_{z}^{(n)}dW_{t}}{\sigma_{z}^{(n)}}\right]^{2}|\hat{y}_{0:T},\sigma_{z}^{(n)}\right]
\]
 with the expectation over the posterior hidden states given the current
$\sigma_{z}^{(n)}$.

The update (\ref{eq:noise-update-rule}) allows to bootstrap APIS
without fine-tuning the noise level for each time series individually,
while at the same time ensures that we learn the best possible control
signal that maximizes the likelihood and maintains a predefined level
of sampling efficiency $\gamma_{\sigma}$. Whenever the effective
sample size is above $\gamma_{\sigma}$, we update $\sigma_{z}$ with
$\eta\simeq0.001$. A small learning rate hinders a sudden drop in
the sampling efficiency and maintains reliable estimates of the controller
and the gradients.

\section{Results}

\subsection{Estimating the Hemodynamic Parameters $\tau_{f}$ and $\tau_{0}$ }

\begin{figure}[!h]
\begin{centering}
\includegraphics[scale=0.5]{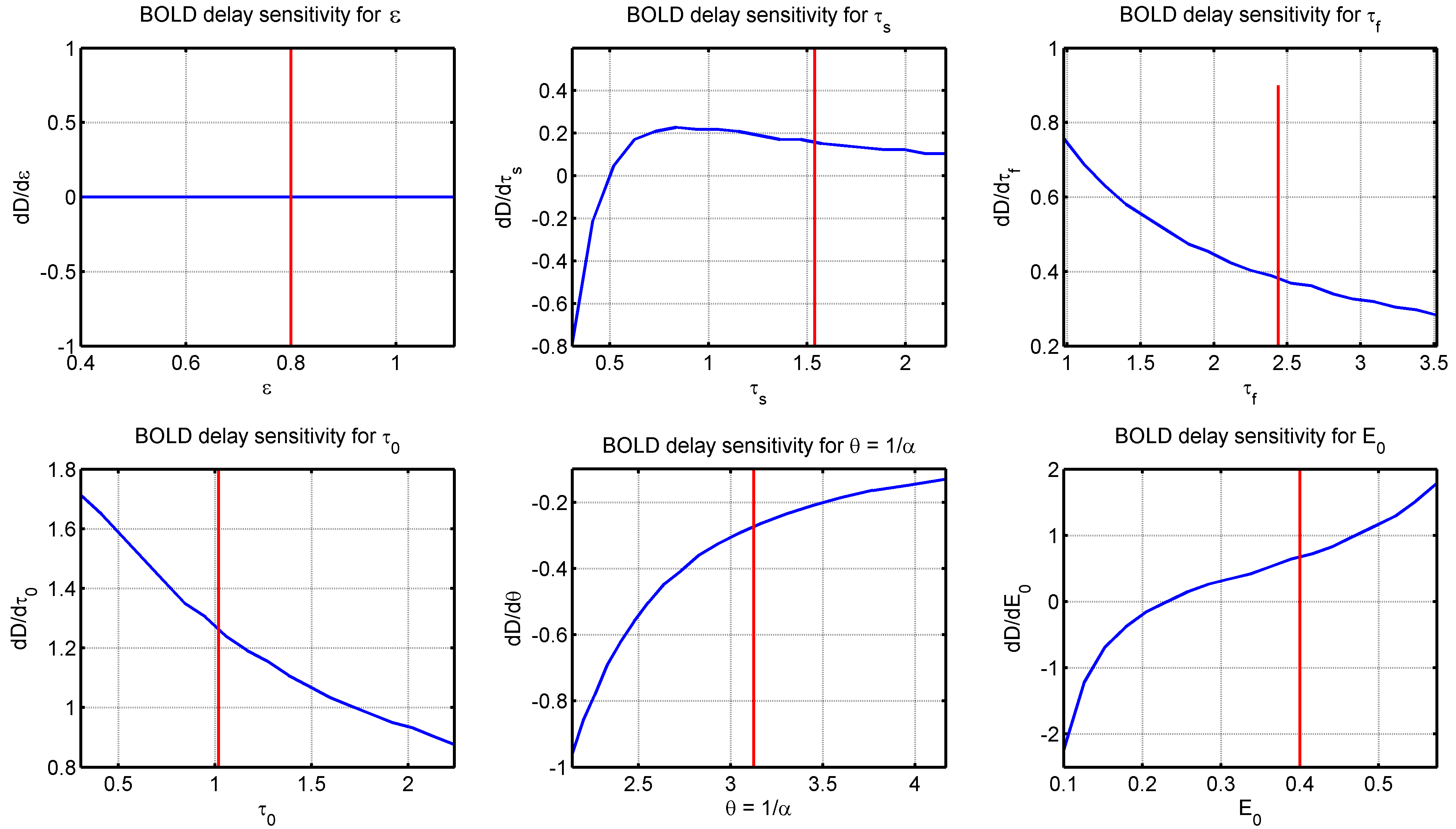}
\par\end{centering}

\caption{Delay sensitivity analysis for the peak-to-peak time (or delay) $D$
between neuronal activity and BOLD signal for the deterministic model
(\ref{eq:neuronal-activity-dyn})-(\ref{eq:BOLD-signal}) with $\sigma_{z}=0$
and $\sigma_{y}=0$. For each panel, we vary the corresponding parameter
keeping all other parameters fixed to typical values. The blue line
indicates the sensitivity $dD/d\omega_{i}$ of the delay for changes
in the corresponding parameter $\omega_{i}\in\Omega=(\epsilon,\tau_{s},\tau_{f},\tau_{0},\alpha,E_{0})$.
The red vertical lines are the typical values in table \tabref{Parameters-of-the-BOLD-trafo}
and $\tau_{0}=1.02$, $\tau_{f}=2.44$. \label{fig:Delay-Analysis}}
\end{figure}

Given an input, the hemodynamic parameters $\Omega=(\epsilon,\tau_{s},\tau_{f},\tau_{0},\alpha,E_{0})$
determine the delay between the neuronal activity and the BOLD signal.
Similarly, when inverting the system to infer the hidden states and
input, the peak of the mean neuronal activity is determined by the
hemodynamic parameters. Since the timing of the neuronal peak is a
proxy for the input timing, estimating the parameters of the hemodynamic
system is crucial.

However, not all parameters have a significant influence on the delay
$D=t_{max,B}-t_{max,z}$ defined as the difference between the timing
of the peak in the BOLD and neuronal signal $z_{t}$. Given an input
to the neuronal system, this influence can be measured by the change
in $D$ when a single parameter is varied. To show this influence,
we compute--by numerical integration--the sensitivity $dD/d\omega_{i}$
of the delay to changes in the parameter $\omega_{i}\in\Omega$. 

Figure \figref{Delay-Analysis} shows this sensitivity for each parameter,
keeping all other fixed to the typical values in table \tabref{Parameters-of-the-BOLD-trafo}
together with $\tau_{0}=1.02$, $\tau_{f}=2.44$. It is observed that
the neuronal efficacy $\epsilon$ has no influence on the delay of
the BOLD amplitude and that, in the neighborhood of the typical values,
the transit time $\tau_{0}$ has the largest sensitivity (lower left
panel), followed by the resting oxygen extraction $E_{0}$ and the
auto-regulation $\tau_{f}$. 

We can use the forward sensitivity method to estimate the values of
hemodynamic parameters given the input and fMRI time series. For simplicity,
only two of the three most relevant parameters are considered, namely
$\tau_{0}$ and $\tau_{f}$. It turns out that including $E_{0}$
in the learning procedure results in local minima if the initialization
is far from the ground truth value of $E_{0}$. This could be an
effect of the insensitivity of the system's output to different parameter
sets, as explored in \citep{deneux2006using}. Hence, the resting
oxygen extraction $E_{0}$ as well as all other parameters will be
fixed to their typical values.

\begin{figure*}[!t]
\begin{centering}
\includegraphics[scale=0.2]{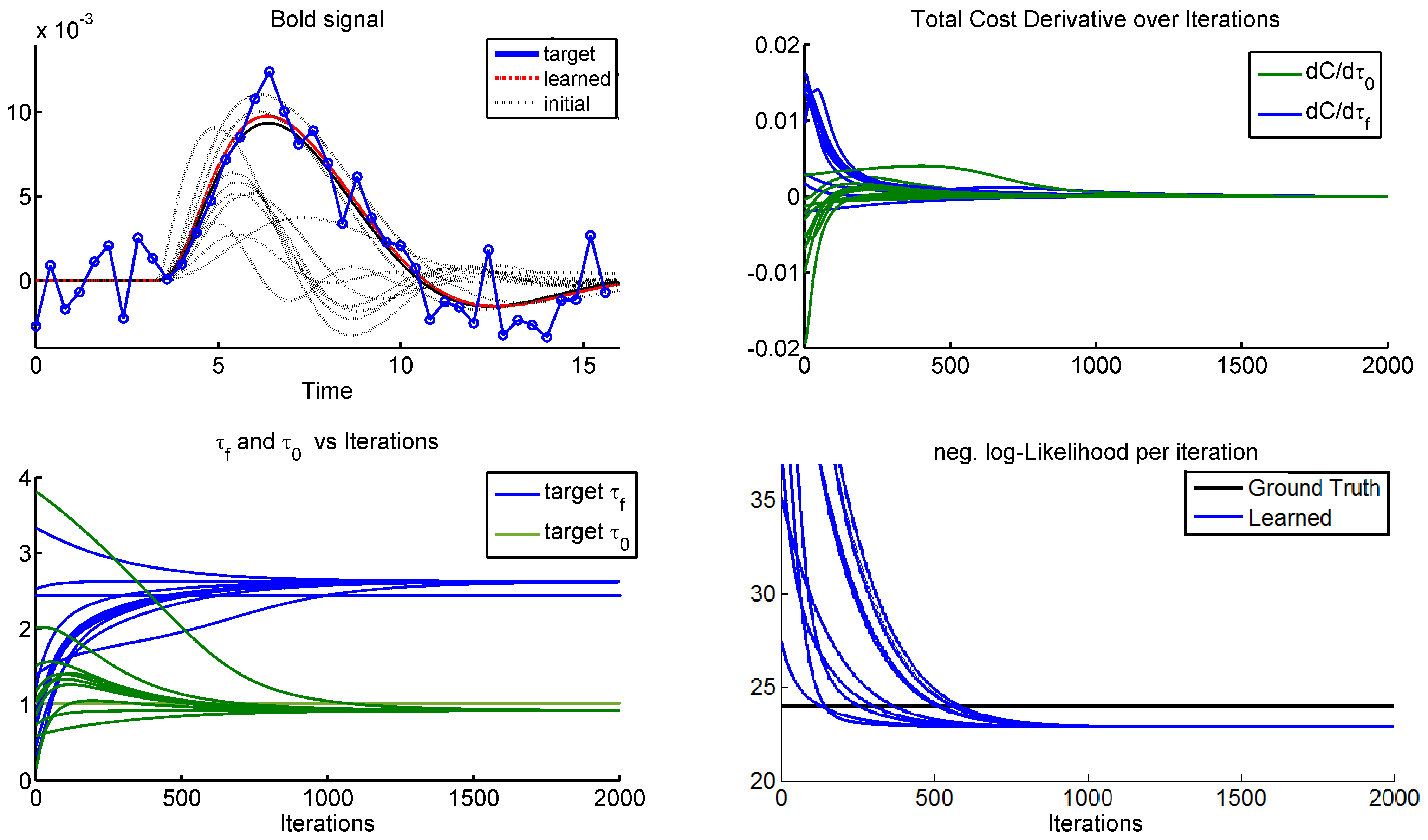}
\par\end{centering}

\caption{Forward sensitivity learning: sparse noisy observations (Top-Left).
The blue markers represent the data sampled. The interval between
observations is $TR=0.4$. Notice that, although the ML estimates
have some error due to the noisy sparse data, the ML solutions are
reliable across a wide range of initializations of the parameters.\label{fig:Forward-sensitivity-learning} }
\end{figure*}

In what follows, the forward sensitivity method for $\tau_{f}$ and
$\tau_{0}$ is analyzed using synthetic data generated as described
in section \subref{Forward-Sensitivity-Equations}. In figure \figref{Forward-sensitivity-learning},
the parameters are estimated from 10 random initializations. As expected,
the estimated parameters are not exactly those of the ground truth,
but all initializations converge to the same values close to them
(left panels). These parameters minimize the neg. log-likelihood of
the data from up to 100 to a value of 23, below that of 24 for the
ground truth (in black, bottom right panel). 

In case there is enough information in the data--either from precise
observations ($\sigma_{obs}=0$) or when the signal is corrupted by
noise but observed at very high time resolution ($TR=dt$)--the parameters
converge to the ground truth. Consequently, the error between the
learned and the ground truth signal is less than $3\times10^{-8}$
over the entire time interval (not shown).

Finally, the parameters $\tau_{0}$ and $\tau_{f}$ are estimated
for each ROI using the mean time series of the first trial. The results
are summarized in table \ref{tab:Estimated-values-of-hemodyn-params}.
In the analysis using synthetic data, we observe estimates that are
close to the ground truth value for small observation noise (not shown).
Given that the mean signal over the first trial has little noise,
it is expected that learned parameters result in a small error in
the input timing of the signals estimated from the second trial.

\begin{table}
\begin{centering}
\begin{tabular}{|c|c|c|c|}
\hline 
\multirow{1}{*}{} & A & V & M\tabularnewline
\hline 
\hline 
$\tau_{0}$ & 1.75 & 1.73 & 2.00\tabularnewline
\hline 
$\tau_{f}$ & 1.56 & 2.58 & 1.32\tabularnewline
\hline 
\end{tabular}
\par\end{centering}

\caption{Estimated values of hemodynamic parameters $\tau_{0}$ and $\tau_{f}$.
These parameters account for most of the variability in the peak-to-peak
delay between the neuronal response and the BOLD signal. The mean
signal of the first trial is used to estimate these parameters, which
are used in the input timing estimates of the second trial. \label{tab:Estimated-values-of-hemodyn-params}}
\end{table}

\subsection{Nonlinear Deconvolution of fMRI Data}

\subsubsection{Hidden Neuronal Activity}

\begin{figure}
\begin{centering}
\includegraphics[scale=0.7]{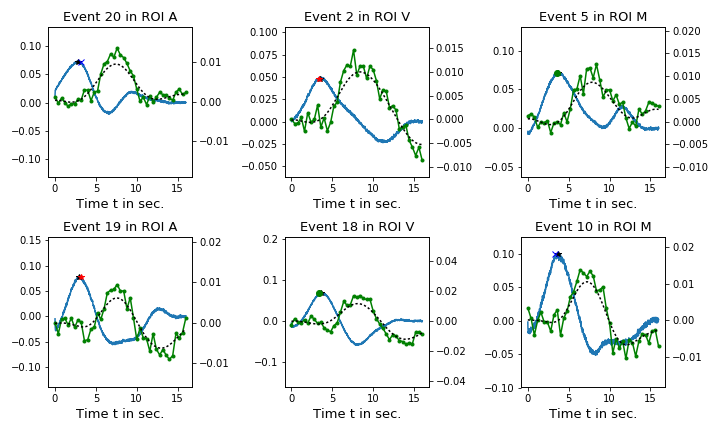}
\par\end{centering}

\caption{Examples of posterior estimates of neuronal activity (blue) and mean
BOLD signal (dashed black). Left axis gives the scale of the neuronal
activity $z_{t}$ and the right the scale of the BOLD signal change.
The green graphs represent the fMRI time series. The black star indicates
the true event timing, while the other markers indicate the estimated
input timing from the maximum neuronal activity. The markers are coded
according to the stimulus modality: blue cross for auditory (A) stimulus,
green dot for visual (V) stimulus and red star for audio-visual stimulus
(AV). \label{fig:Examples-posterior-estimates}}
\end{figure}

APIS is applied individually to the fMRI time series of the second
trial. For all estimations, $5\times10^{4}$ particles are used over
120 iterations to infer the neuronal activity. Unless stated otherwise
all simulations use the typical values in table \tabref{Parameters-of-the-BOLD-trafo},
the estimated values in table \tabref{Estimated-values-of-hemodyn-params}
and the noise level is initialized at $\sigma_{z}=0.3$.

Figure \figref{Examples-posterior-estimates} shows typical examples
of the posterior estimates using APIS. Each column represents a ROI.
The markers show the stimulus time (black star) and the estimated
input time. The color of the markers represent the modality of the
stimulus presented (A: blue, V: green, AV: red). The neuronal signals
(blue line) follow very closely the estimated input due to the short
time scale of the system. In addition, they have large amplitudes
and a clear peak around the input timing. This gives BOLD signals
with high likelihood (black dashed lines). 

\begin{figure}
\begin{centering}
\includegraphics[scale=1.2]{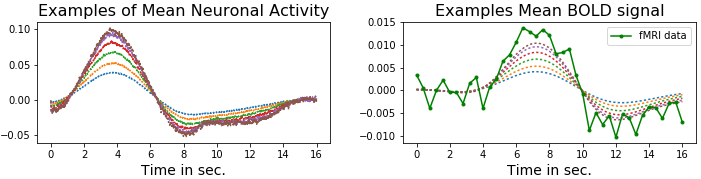}
\par\end{centering}

\caption{Examples of posterior mean estimates depending on the noise level:
from the blue solution with $\sigma_{dyn}=0.44$ to the brown solution
with $\sigma_{dyn}=1.17$. The higher the noise $\sigma_{dyn}$, the
stronger the inferred input and, thus, the neuronal signal has a larger
amplitude (left panel). This gives higher amplitudes in the corresponding
mean BOLD signals (right). Nevertheless, the width of the resulting
neuronal signals does not change significantly. \label{fig:Examples-of-posterior}}
\end{figure}

The amplitudes of both, the neuronal and the BOLD signals, depend
directly on the dynamic noise $\sigma_{dyn}$. In figure \figref{Examples-of-posterior},
examples of the posterior mean estimates of both the neuronal activity
and the BOLD signal are shown for a given time series (green lines
with markers). The noise $\sigma_{dyn}$ varies from $0.44-1.17$,
which result in different mean signals with varying amplitudes, from
low to high respectively. 

While the maximum of the neuronal activity is a good proxy for the
input timing, the time scale of the true input cannot be captured
by the inferred signal $u(z_{t},t)$. This is due to the well known
low pass-filter characteristic of the hemodynamic transformation,
which makes fast variations in the neuronal activity have a minor
impact on the resulting BOLD signal. Hence, although the amplitude
of the inferred input increases with the noise level, its width of
several seconds remains broad regardless of the noise, see left panel
on figure \figref{Examples-of-posterior}. These results are robust
against changes in the model to have faster time scales, for instance
a geometric Brownian motion with a non vanishing diffusion term at
the origin, or changes in the time scale $A$ of the neuronal system.

The poor temporal resolution of the neuronal signal resulting from
inverting the hemodynamic system is confirmed with a random grid search
over the neuronal space at each time point $z_{t}$ to obtain ML-estimates.
Using this method the resulting signal is similar to our results;
it has also a peak centered around the input and its width is several
seconds (not shown). Hence, without explicitly modeling fast changes
in the input over time, the deconvolved signal will not reflect fast
fluctuations of the inputs.

\subsubsection{Comparison with Bootstrap Particle Filter-Smoother}

\begin{figure}
\begin{centering}
\includegraphics[width=0.8\textwidth]{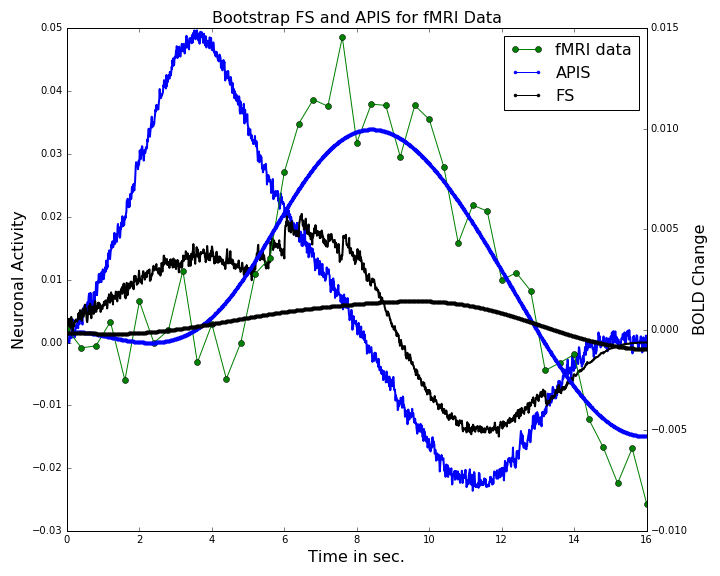}
\par\end{centering}

\caption{Comparison of APIS vs the vanilla bootstrap filter-smoother (FS).
The left y-axis shows the strength of the neuronal activity $z_{t}$,
while the right axis the BOLD response change. The jittered lines
are the neuronal activity estimates of APIS (blue) and FS (black).
The thick continuous lines are their corresponding BOLD response.
Notice the low amplitude of the FS estimates. This results in a negative
log-likelihood score of 374; much higher than the score for APIS with
97. \label{fig:Comparison-of-APIS-vs-FS}}
\end{figure}

In what follows we compare APIS to the vanilla flavor bootstrap particle
filter-smoother (FS). The algorithm for FS is implemented as in \citet{lindsten2013backward}.
We estimate the smoothing distribution over the neuronal activity
using 40 workers with $5000$ particles each. The computation of the
statistics of the posterior is parallelized across all workers. Since
the effective samples of the FS deteriorates for early times, the
variance of the estimates at these times is large. For this reason,
100 forward passes on each CPU are performed to obtain better estimates.
The noise level $\sigma_{dyn}$ is set to the same value (0.39) as
the one estimated by the EM procedure for APIS. 

Figure \ref{fig:Comparison-of-APIS-vs-FS} compares the estimates
from APIS and FS for event \#2 in the V ROI. The result shows that
the FS has problems estimating a large, clear amplitude of the neuronal
activity due to the strong model mismatch from the lack of input.
In contrast, the control drift $u(x_{t},t)$ in APIS accounts for
the lack of input in our model, giving a clear peak in the neuronal
activity centered at the event time. Scoring both results in terms
of the negative log-likelihood, the FS has a much higher score of
374 compared to APIS with 97. Hence, the FS samples represent poorly
the data and are bias towards the prior stationary distribution given
by the uncontrolled dynamics.

\subsubsection{Validation via Input Timing Estimates from Single Event fMRI Time
Series}

\begin{figure*}[!t]
\begin{centering}
\includegraphics[width=0.95\textwidth]{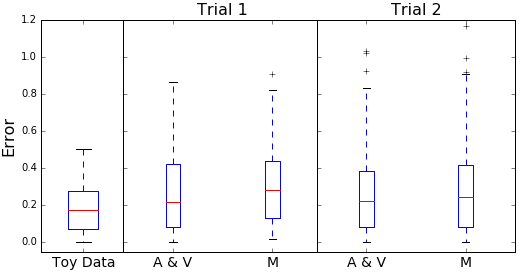}
\par\end{centering}

\caption{Performance summary of timing estimates for APIS. Synthetic data is
used to define a reference empirical distribution of timing errors
(left box plot). The hemodynamic model is estimated using trial 1
and trial 2 is used to evaluate the generalization of the deconvolution
to unseen data. Although the variation in the inter-quartile range
is up to 62\% higher than that of the reference distribution, over
75\% of the error in both trials lie within the range of the reference
distribution and all third quartiles lie around $TR=0.4$. The median
values lie within the third quartile of the reference distribution.
Notice that the models estimated in trial 1 generalize well in trial
2. \label{fig:Error-of-timing}}
\end{figure*}

As an indication of the performance of the proposed deconvolution,
we consider box plots representing the empirical distribution over
errors between the estimated timings and the actual event time. Comparing
these distributions with the \textquotedbl{}reference\textquotedbl{}
empirical distribution obtained from synthetic data gives us an idea
of how well the deconvolution procedure, together with the parameter
estimates, extracts timing information at the neuronal level. 

For the error distribution from synthetic data, we generate 30 time
series with the same characteristics of the real data (TR, BOLD change
amplitude, trial length, etc) using the same ground truth as in \subref{Forward-Sensitivity-Equations}
but with a neuronal noise of $\sigma_{z}=0.01$. To infer the neuronal
signals, the same algorithmic parameters are used in synthetic and
real data.

Figure \figref{Error-of-timing} shows the box plots for the synthetic
data, the combined errors for the sensory ROIs and the errors from
the motor ROI in each trial. We consider the sensory regions and the
motor region separately because the inputs to the sensory ROIs are
exactly the visual and auditory cues, which are known with precision,
while the inputs to the motor ROI are less certain. The first box
plot on the left side shows that even in the idealized case of having
the exact hemodynamic model, the estimated timings have a median error
of $0.18$ with an inter-quartile range of $[0.08,0.28]$. This is
to be expected due to the noise in the system. The median error in
both trials lies between $[0.22,0.28]$ and thus within the third
quartile of the reference distribution. Interestingly, the median
of the sensory ROIs have very similar values about 23\% higher than
that of the reference, but the median of the motor ROI is in both
trials up to 50\% higher. Nevertheless, in all cases more than 75\%
of the errors lie within the range of the reference distribution and
there is an overall count of only 7 outliers. All third quartile values
in both trials lie around the TR value.

Although in both trials the variation in the error is about 62\% higher
than in the reference distribution, we observe consistency between
the trials. Hence, the parameters estimated in the first trial generalize
well in the second trial. The source of the higher error is probably
due to both, a bias in the estimation of the hemodynamic system and
a large variance in the data, possibly from the effects of the neuronal
noise on the BOLD response, as this can be observed in synthetic data
(not shown).

These results show that APIS is capable of extracting timing information
with a very high resolution, well beyond the typical time scales of
the hemodynamic transformation and the TR. Even more, comparing the
error obtained from 2 different trials from the same subject--one
used to fit the model--it is possible to assess the generality of
the estimated model.

\section{Discussion}

In this article, the adaptive importance sampler APIS \citep{ruiz2017particle}
is applied to fMRI data obtained during a reaction time experiment
to infer the latent neuronal activity in the visual, auditory and
motor ROIs. In addition, APIS is extended by an EM-based procedure
to increase the neuronal noise in the system such that signals with
high likelihood can be inferred while maintaining the sampler at a
fixed efficiency.  We show that this procedure is capable of compensating
for signals not included in the model to obtain accurate results.
In this case, the input to the ROI is disregarded in the reconstruction
of the latent neuronal activity and the aim is to obtain accurate
absolute timing estimates of the events as a way of validating the
method on real fMRI time series. Accuracy is measured in terms of
the empirical distribution of the error in input timing estimates
compared to a reference distribution obtained from synthetic data
given the ground truth model. In addition, our method shows a clear
advantage over particle methods in terms of efficiently minimizing
the variance of the neural estimates and the negative likelihood.

In some cases, the ML-estimates of the hemodynamic parameters resulted
in a higher MSE in the event timing estimates of trial 2. One source
of error is a bias in the model caused by a distortion of the mean
signal used in training due to outliers in the data. To avoid this
effect, it might be wise to consider neuronal noise in the model estimation
procedure and obtain ML-estimates for each time series individually.
For this, APIS can be merged readily with the Forward Sensitivity
method to jointly estimate parameters and hidden states.

Another source of bias in the timing estimates can be caused by a
fine-tuning problem where other hemodynamic parameters must be learned.
Since the application of the Forward Sensitivity method is cumbersome
for larger number of parameters, other possible modifications of the
joint estimation procedure exist. For instance, the addition of a
small noise signal to the hemodynamic system allows the implementation
of the EM algorithm on the full system. In this case, the parameters
are updated with the gradient of the state transition density under
the posterior statistics. This gives simpler learning rules than the
Forward Sensitive method, but it requires the estimation of a control
signal for each additional noisy degree of freedom. Nonetheless, APIS
has the additional benefit that non-neuronal sources in the hemodynamic
response could be detected in a fully controlled setting because,
as shown here, hidden signals that are not modeled can be captured
by the controller. An additional advantage, albeit the higher computational
effort required, is the better behavior of the effective sample size,
which is higher and less sensible to changes in the neuronal noise.
In this work, however, we consider only a 1-dimensional linear feedback
controller for simplicity.

Interesting parallels to our case study are found in \citep{havlicek2011dynamic}
and more recently \citep{sreenivasan2015nonparametric}. In the former,
toy data was used to analyze their proposed method for short stimuli
of about $200\ ms$. Interestingly, the cubature Kalman Filter-Smoother
(SCKS) presented there obtained similar broad neuronal signals, but
there was no report on the accuracy of the input timing estimates.
In the latter article, a non-parametric approach was used to asses
the susceptibility of the SCKS to over-fit the data. Both methods
were applied to fMRI time series obtained during the presentation
of visual stimuli with a duration of 2 s. In this case too, the SCKS
estimates had clear peaks in the neuronal response around the stimulus
timing, but the stimulus duration was one order of magnitude larger
than in our case study and there was no report on precise input timing
estimates. Hence, these methods were tested in a different dynamic
regime of the hemodynamic transformation than here. On the contrary,
in our case study both aspects--short stimuli and real fMRI data--are
combined.

The lack of direct measurements of the neuronal activity makes it
difficult to asses the quality of any estimation, but the theoretical
derivation of APIS ensures the optimality of the posterior estimates
given the model and a sufficiently high number of effective samples.
Although we recognize the need to compare our proposed method with
interesting alternatives, a fair comparison is difficult because the
accuracy of the different methods in different dynamic regimes is
not known and there is no ground truth available. Thus, it is important
to clarify which of the available methods are good approximations
for the different stimuli modalities and dynamic regimes of the nonlinear
hemodynamic system, and how they compare to our adaptive importance
sampling method, which is applicable in any dynamic regime. Hence,
finding good benchmarks for all methods with many types of stimulus
modalities is urgently needed and an extensive comparison of alternative
methods should be addressed in future work. In this analysis, the
question of how accurate the Gaussian approximations are in the different
dynamic regimes would require special attention. 

To overcome the lack of a ground truth, the proposed method was validated
by its ability to directly extract timing information out of the time
series. Inferring the timing of a stimulus shorter than the TR of
the acquisition requires very precise estimations of the neuronal
activity, especially if very fast changes are suppressed by the hemodynamic
transformation and the only retrievable information are signals with
a temporal resolution an order of magnitude larger. Hence, we believe
that the proposed analysis can serve as a benchmark on the precision
of timing information retrieval for deconvolution methods of fMRI
time series.

Finally, the proposed method can accommodate any nonlinear model that
can be formulated as a stochastic differential equation, for instance
the recent model \citep{havlicek2015physiologically}. Moreover, the
model can be easily extended to accommodate context dependent connectivity
and other confound signals like heart rate or head motion. This generality
and flexibility together with the demonstrated accuracy and efficiency
makes the proposed method an interesting alternative to other nonlinear
deconvolution methods for event related and resting state fMRI data.

\subsubsection*{Acknowledgments}

The work of HC Ruiz Euler was supported by the European Commission
through the FP7 Marie Curie Initial Training Network 289146, NETT:
Neural Engineering Transformative Technologies.

\pagebreak{}

\bibliographystyle{plainnat}
\bibliography{C:/Users/HCRuiz/SURFdrive/PhD_backup/References/PhD_Bib_Database}

\begin{thebibliography}{41}
\providecommand{\natexlab}[1]{#1}
\providecommand{\url}[1]{\texttt{#1}}
\expandafter\ifx\csname urlstyle\endcsname\relax
  \providecommand{\doi}[1]{doi: #1}\else
  \providecommand{\doi}{doi: \begingroup \urlstyle{rm}\Url}\fi

\bibitem[Birn et~al.(2001)Birn, Saad, and Bandettini]{birn2001spatial}
Rasmus~M Birn, Ziad~S Saad, and Peter~A Bandettini.
\newblock Spatial heterogeneity of the nonlinear dynamics in the fmri bold
  response.
\newblock \emph{Neuroimage}, 14\penalty0 (4):\penalty0 817--826, 2001.

\bibitem[Briers et~al.(2010)Briers, Doucet, and Maskell]{briers2010smoothing}
Mark Briers, Arnaud Doucet, and Simon Maskell.
\newblock Smoothing algorithms for state--space models.
\newblock \emph{Annals of the Institute of Statistical Mathematics},
  62\penalty0 (1):\penalty0 61--89, 2010.

\bibitem[Buxton et~al.(1998)Buxton, Wong, and Frank]{buxton1998dynamics}
Richard~B Buxton, Eric~C Wong, and Lawrence~R Frank.
\newblock Dynamics of blood flow and oxygenation changes during brain
  activation: the balloon model.
\newblock \emph{Magnetic resonance in medicine}, 39\penalty0 (6):\penalty0
  855--864, 1998.

\bibitem[Daunizeau et~al.(2009)Daunizeau, Friston, and
  Kiebel]{daunizeau2009variational}
Jean Daunizeau, Karl~J Friston, and Stefan~J Kiebel.
\newblock Variational bayesian identification and prediction of stochastic
  nonlinear dynamic causal models.
\newblock \emph{Physica D: Nonlinear Phenomena}, 238\penalty0 (21):\penalty0
  2089--2118, 2009.

\bibitem[David et~al.(2008)David, Guillemain, Saillet, Reyt, Deransart,
  Segebarth, and Depaulis]{david2008identifying}
Olivier David, Isabelle Guillemain, Sandrine Saillet, Sebastien Reyt, Colin
  Deransart, Christoph Segebarth, and Antoine Depaulis.
\newblock Identifying neural drivers with functional mri: an
  electrophysiological validation.
\newblock \emph{PLoS biology}, 6\penalty0 (12):\penalty0 e315, 2008.

\bibitem[Deneux and Faugeras(2006)]{deneux2006using}
Thomas Deneux and Olivier Faugeras.
\newblock Using nonlinear models in fmri data analysis: model selection and
  activation detection.
\newblock \emph{NeuroImage}, 32\penalty0 (4):\penalty0 1669--1689, 2006.

\bibitem[Douc et~al.(2011)Douc, Garivier, Moulines, and Olsson]{Douc2011}
Randal Douc, Aur\'elien Garivier, Eric Moulines, and Jimmy Olsson.
\newblock Sequential monte carlo smoothing for general state space hidden
  markov models.
\newblock \emph{The Annals of Applied Probability}, 21\penalty0 (6):\penalty0
  2109--2145, Dec 2011.
\newblock ISSN 1050-5164.
\newblock \doi{10.1214/10-AAP735}.
\newblock URL \url{http://projecteuclid.org/euclid.aoap/1322057317}.

\bibitem[Fleming and Mitter(1982)]{fleming1982optimal}
Wendell~H Fleming and Sanjoy~K Mitter.
\newblock Optimal control and nonlinear filtering for nondegenerate diffusion
  processes.
\newblock \emph{Stochastics: An International Journal of Probability and
  Stochastic Processes}, 8\penalty0 (1):\penalty0 63--77, 1982.

\bibitem[Formisano and Goebel(2003)]{formisano2003tracking}
Elia Formisano and Rainer Goebel.
\newblock Tracking cognitive processes with functional mri mental chronometry.
\newblock \emph{Current opinion in neurobiology}, 13\penalty0 (2):\penalty0
  174--181, 2003.

\bibitem[Friston et~al.(2010)Friston, Stephan, Li, and
  Daunizeau]{friston2010generalised}
Karl Friston, Klaas Stephan, Baojuan Li, and Jean Daunizeau.
\newblock Generalised filtering.
\newblock \emph{Mathematical Problems in Engineering}, 2010, 2010.

\bibitem[Friston(2008)]{friston2008avariational}
Karl~J Friston.
\newblock Variational filtering.
\newblock \emph{NeuroImage}, 41\penalty0 (3):\penalty0 747--766, 2008.

\bibitem[Friston et~al.(1998{\natexlab{a}})Friston, Fletcher, Josephs, Holmes,
  Rugg, and Turner]{friston1998event}
Karl~J Friston, P~Fletcher, Oliver Josephs, ANDREW Holmes, MD~Rugg, and Robert
  Turner.
\newblock Event-related fmri: characterizing differential responses.
\newblock \emph{Neuroimage}, 7\penalty0 (1):\penalty0 30--40,
  1998{\natexlab{a}}.

\bibitem[Friston et~al.(1998{\natexlab{b}})Friston, Josephs, Rees, and
  Turner]{friston1998nonlinear}
Karl~J Friston, Oliver Josephs, Geraint Rees, and Robert Turner.
\newblock Nonlinear event-related responses in fmri.
\newblock \emph{Magnetic resonance in medicine}, 39\penalty0 (1):\penalty0
  41--52, 1998{\natexlab{b}}.

\bibitem[Friston et~al.(2000)Friston, Mechelli, Turner, and
  Price]{friston2000nonlinear}
Karl~J Friston, Andrea Mechelli, Robert Turner, and Cathy~J Price.
\newblock Nonlinear responses in fmri: the balloon model, volterra kernels, and
  other hemodynamics.
\newblock \emph{NeuroImage}, 12\penalty0 (4):\penalty0 466--477, 2000.

\bibitem[Friston et~al.(2003)Friston, Harrison, and Penny]{friston2003dynamic}
Karl~J Friston, Lee Harrison, and Will Penny.
\newblock Dynamic causal modelling.
\newblock \emph{Neuroimage}, 19\penalty0 (4):\penalty0 1273--1302, 2003.

\bibitem[Friston et~al.(2008)Friston, Trujillo-Barreto, and
  Daunizeau]{friston2008bvariational}
Karl~J Friston, N~Trujillo-Barreto, and Jean Daunizeau.
\newblock Dem: a variational treatment of dynamic systems.
\newblock \emph{Neuroimage}, 41\penalty0 (3):\penalty0 849--885, 2008.

\bibitem[Griffanti et~al.(2017)Griffanti, Douaud, Bijsterbosch, Evangelisti,
  Alfaro-Almagro, Glasser, Duff, Fitzgibbon, Westphal, Carone,
  et~al.]{griffanti2017hand}
Ludovica Griffanti, Gwena{\"e}lle Douaud, Janine Bijsterbosch, Stefania
  Evangelisti, Fidel Alfaro-Almagro, Matthew~F Glasser, Eugene~P Duff, Sean
  Fitzgibbon, Robert Westphal, Davide Carone, et~al.
\newblock Hand classification of fmri ica noise components.
\newblock \emph{NeuroImage}, 154:\penalty0 188--205, 2017.

\bibitem[Havlicek et~al.(2011)Havlicek, Friston, Jan, Brazdil, and
  Calhoun]{havlicek2011dynamic}
Martin Havlicek, Karl~J Friston, Jiri Jan, Milan Brazdil, and Vince~D Calhoun.
\newblock Dynamic modeling of neuronal responses in fmri using cubature kalman
  filtering.
\newblock \emph{NeuroImage}, 56\penalty0 (4):\penalty0 2109--2128, 2011.

\bibitem[Havlicek et~al.(2015)Havlicek, Roebroeck, Friston, Gardumi, Ivanov,
  and Uludag]{havlicek2015physiologically}
Martin Havlicek, Alard Roebroeck, Karl Friston, Anna Gardumi, Dimo Ivanov, and
  Kamil Uludag.
\newblock Physiologically informed dynamic causal modeling of fmri data.
\newblock \emph{NeuroImage}, 122:\penalty0 355--372, 2015.

\bibitem[Jimenez and Ozaki(2003)]{jimenez2003local}
JC~Jimenez and T~Ozaki.
\newblock Local linearization filters for non-linear continuous-discrete state
  space models with multiplicative noise.
\newblock \emph{International Journal of Control}, 76\penalty0 (12):\penalty0
  1159--1170, 2003.

\bibitem[Johnston et~al.(2008)Johnston, Duff, Mareels, and
  Egan]{johnston2008nonlinear}
Leigh~A Johnston, Eugene Duff, Iven Mareels, and Gary~F Egan.
\newblock Nonlinear estimation of the bold signal.
\newblock \emph{NeuroImage}, 40\penalty0 (2):\penalty0 504--514, 2008.

\bibitem[Kappen(2005)]{kappen2005linear}
Hilbert~J Kappen.
\newblock Linear theory for control of nonlinear stochastic systems.
\newblock \emph{Physical review letters}, 95\penalty0 (20):\penalty0 200201,
  2005.

\bibitem[Kappen(2011)]{kappen2011optimal}
Hilbert~J Kappen.
\newblock Optimal control theory and the linear bellman equation.
\newblock \emph{Inference and Learning in Dynamic Models}, pages 363--387,
  2011.

\bibitem[Kappen et~al.(2012)Kappen, G{\'o}mez, and Opper]{kappen2012optimal}
Hilbert~J Kappen, Vicen{\c{c}} G{\'o}mez, and Manfred Opper.
\newblock Optimal control as a graphical model inference problem.
\newblock \emph{Machine learning}, 87\penalty0 (2):\penalty0 159--182, 2012.

\bibitem[Kappen and Ruiz(2016)]{kappen2016adaptive}
Hilbert~Johan Kappen and Hans~Christian Ruiz.
\newblock Adaptive importance sampling for control and inference.
\newblock \emph{Journal of Statistical Physics}, 162\penalty0 (5):\penalty0
  1244--1266, 2016.

\bibitem[Katwal et~al.(2013)Katwal, Gore, Gatenby, and
  Rogers]{katwal2013measuring}
Santosh~B Katwal, John~C Gore, J~Christopher Gatenby, and Baxter~P Rogers.
\newblock Measuring relative timings of brain activities using fmri.
\newblock \emph{NeuroImage}, 66:\penalty0 436--448, 2013.

\bibitem[Li et~al.(2011)Li, Daunizeau, Stephan, Penny, Hu, and
  Friston]{li2011generalised}
Baojuan Li, Jean Daunizeau, Klaas~E Stephan, Will Penny, Dewen Hu, and Karl
  Friston.
\newblock Generalised filtering and stochastic dcm for fmri.
\newblock \emph{neuroimage}, 58\penalty0 (2):\penalty0 442--457, 2011.

\bibitem[Liao et~al.(2002)Liao, Worsley, Poline, Aston, Duncan, and
  Evans]{liao2002estimating}
Chien~Heng Liao, Keith~J Worsley, J-B Poline, John~AD Aston, Gary~H Duncan, and
  Alan~C Evans.
\newblock Estimating the delay of the fmri response.
\newblock \emph{NeuroImage}, 16\penalty0 (3):\penalty0 593--606, 2002.

\bibitem[Lindquist(2008)]{lindquist2008statistical}
Martin~A Lindquist.
\newblock The statistical analysis of fmri data.
\newblock \emph{Statistical Science}, pages 439--464, 2008.

\bibitem[Lindsten and Sch{\"o}n(2013)]{lindsten2013backward}
Fredrik Lindsten and Thomas~B Sch{\"o}n.
\newblock Backward simulation methods for monte carlo statistical inference.
\newblock \emph{Foundations and Trends in Machine Learning}, 6\penalty0
  (1):\penalty0 1--143, 2013.

\bibitem[Miezin et~al.(2000)Miezin, Maccotta, Ollinger, Petersen, and
  Buckner]{miezin2000characterizing}
Francis~M Miezin, L~Maccotta, JM~Ollinger, SE~Petersen, and RL~Buckner.
\newblock Characterizing the hemodynamic response: effects of presentation
  rate, sampling procedure, and the possibility of ordering brain activity
  based on relative timing.
\newblock \emph{Neuroimage}, 11\penalty0 (6):\penalty0 735--759, 2000.

\bibitem[Murray and Storkey(2011)]{murray2011particle}
Lawrence Murray and Amos Storkey.
\newblock Particle smoothing in continuous time: A fast approach via density
  estimation.
\newblock \emph{Signal Processing, IEEE Transactions on}, 59\penalty0
  (3):\penalty0 1017--1026, 2011.

\bibitem[Narsude et~al.(2015)Narsude, Gallichan, Van Der~Zwaag, Gruetter, and
  Marques]{narsude2015three}
Mayur Narsude, Daniel Gallichan, Wietske Van Der~Zwaag, Rolf Gruetter, and
  Jos{\'e}~P Marques.
\newblock Three-dimensional echo planar imaging with controlled aliasing: A
  sequence for high temporal resolution functional mri.
\newblock \emph{Magnetic resonance in medicine}, 2015.

\bibitem[Riera et~al.(2004)Riera, Watanabe, Kazuki, Naoki, Aubert, Ozaki, and
  Kawashima]{riera2004state}
Jorge~J Riera, Jobu Watanabe, Iwata Kazuki, Miura Naoki, Eduardo Aubert, Tohru
  Ozaki, and Ryuta Kawashima.
\newblock A state-space model of the hemodynamic approach: nonlinear filtering
  of bold signals.
\newblock \emph{NeuroImage}, 21\penalty0 (2):\penalty0 547--567, 2004.

\bibitem[Roweis and Ghahramani(2001)]{roweis2001learning}
Sam Roweis and Zoubin Ghahramani.
\newblock Learning nonlinear dynamical systems using the
  expectation--maximization algorithm.
\newblock \emph{Kalman filtering and neural networks}, 6:\penalty0 175--220,
  2001.

\bibitem[Ruiz and Kappen(2017)]{ruiz2017particle}
Hans-Christian Ruiz and Hilbert~J Kappen.
\newblock Particle smoothing for hidden diffusion processes: Adaptive path
  integral smoother.
\newblock \emph{IEEE Transactions on Signal Processing}, 65\penalty0
  (12):\penalty0 3191--3203, 2017.

\bibitem[Sengupta et~al.(2014)Sengupta, Friston, and
  Penny]{sengupta2014efficient}
Biswa Sengupta, Karl~J Friston, and William~D Penny.
\newblock Efficient gradient computation for dynamical models.
\newblock \emph{NeuroImage}, 98:\penalty0 521--527, 2014.

\bibitem[Smith et~al.(2011)Smith, Miller, Salimi-Khorshidi, Webster, Beckmann,
  Nichols, Ramsey, and Woolrich]{smith2011network}
Stephen~M Smith, Karla~L Miller, Gholamreza Salimi-Khorshidi, Matthew Webster,
  Christian~F Beckmann, Thomas~E Nichols, Joseph~D Ramsey, and Mark~W Woolrich.
\newblock Network modelling methods for fmri.
\newblock \emph{Neuroimage}, 54\penalty0 (2):\penalty0 875--891, 2011.

\bibitem[Sreenivasan et~al.(2015)Sreenivasan, Havlicek, and
  Deshpande]{sreenivasan2015nonparametric}
Karthik~Ramakrishnan Sreenivasan, Martin Havlicek, and Gopikrishna Deshpande.
\newblock Nonparametric hemodynamic deconvolution of fmri using homomorphic
  filtering.
\newblock \emph{IEEE transactions on medical imaging}, 34\penalty0
  (5):\penalty0 1155--1163, 2015.

\bibitem[Stephan et~al.(2008)Stephan, Kasper, Harrison, Daunizeau, den Ouden,
  Breakspear, and Friston]{stephan2008nonlinear}
Klaas~Enno Stephan, Lars Kasper, Lee~M Harrison, Jean Daunizeau, Hanneke~EM den
  Ouden, Michael Breakspear, and Karl~J Friston.
\newblock Nonlinear dynamic causal models for fmri.
\newblock \emph{Neuroimage}, 42\penalty0 (2):\penalty0 649--662, 2008.

\bibitem[Wager et~al.(2005)Wager, Vazquez, Hernandez, and
  Noll]{wager2005accounting}
Tor~D Wager, Alberto Vazquez, Luis Hernandez, and Douglas~C Noll.
\newblock Accounting for nonlinear bold effects in fmri: parameter estimates
  and a model for prediction in rapid event-related studies.
\newblock \emph{NeuroImage}, 25\penalty0 (1):\penalty0 206--218, 2005.

\end{thebibliography}

\end{document}